\documentclass[prl,superscriptaddress,showpacs,byrevtex,twocolumn]{revtex4}
\usepackage[]{amssymb}
\usepackage[]{graphicx}

\begin{document}

\bibliographystyle{apsrev}

\title{Signature of quantum criticality in
photoemission spectroscopy}

\author{M. Klein}
\affiliation{Universit\"at W\"urzburg, Experimentelle Physik II,
  Am Hubland, D-97074 W\"urzburg, Germany} 
\author{A. Nuber}
\affiliation{Universit\"at W\"urzburg, Experimentelle Physik II,
  Am Hubland, D-97074 W\"urzburg, Germany} 
\author{F. Reinert}
\affiliation{Universit\"at W\"urzburg, Experimentelle Physik II,
  Am Hubland, D-97074 W\"urzburg, Germany} 
\affiliation{Forschungszentrum Karlsruhe, Gemeinschaftslabor f\"ur Nanoanalytik, D-76021 Karlsruhe, Germany}
\author{J. Kroha}
\affiliation{Universit\"at Bonn, Physikalisches Institut,
  Nussallee 12, D-53115 Bonn, Germany} 
\author{O. Stockert}
\affiliation{Max Planck Institute for Chemical Physics of Solids, 
N\"othnitzer Str. 40, 01187 Dresden, Germany}
\author{H. v.\ L\"ohneysen}
\affiliation{Physikalisches Institut,
Universit\"at Karlsruhe (TH), D-76128 Karlsruhe, Germany}
\affiliation{Forschungszentrum Karlsruhe, Institut f\"ur Festk\"operphysik, D-76021 Karlsruhe, Germany}

\date{\today}

\begin{abstract}
A quantum phase transition (QPT) in a heavy-fermion (HF) compound
may destroy the Fermi liquid 
groundstate. However, the conditions for this breakdown have 
remained obscure.  
We report the first direct investigation of heavy quasiparticle formation and 
breakdown in the canonical system CeCu$_{6-x}$Au$_x$ 
by ultraviolet photoemission spectroscopy at elevated temperatures 
without the complications of lattice coherence. 
Surprisingly, the single-ion Kondo energy scale $T_K$ exhibits an abrupt 
step near the quantum critical  
Au concentration of $x_c=0.1$. We show theoretically that this step is
expected from a highly non-linear renormalization of the local spin coupling 
at each Ce site, induced by spin fluctuations on neighboring sites. 
It provides a general high-temperature indicator for HF quasiparticle 
breakdown at a QPT.
\end{abstract}
\maketitle

%Quantum phase transitions (QPT) in heavy-fermion (HF) systems are among the 
%most striking examples for the breakdown of Fermi liquid (FL) behavior and the 
%formation of a new groundstate of matter, as recent experiments indicate 
%\cite{loehneysen07,steglich08}

A quantum phase transition (QPT) is a second-order transition 
occurring at zero temperature, driven by a nonthermal control parameter 
such as composition or pressure. Here, two competing groundstates are 
separated by a quantum critical point (QCP). Heavy-fermion (HF) systems, 
in particular, show striking deviations from Fermi-liquid (FL) behavior 
at a QPC between a magnetically ordered and a paramagnetic phase
\cite{loehneysen07,steglich08}.
In the standard Hertz-Millis (HM) scenario 
\cite{hertz76,millis93} only the bosonic fluctuations 
of the order parameter become quantum critical, i.e., long-ranged in 
space and time, leading to anomalous behavior of physical 
quantities, but leaving the fermionic quasiparticles intact. 
However, in a heavy fermion (HF) system the order-parameter fluctuations 
always couple to the spin degree of freedom of the fermionic excitations, 
so that the latter may become critical as well and disintegrate. 
In this case, the formation of 
the Kondo spin singlet between the conduction electron and the local 
$4f$ magnetic moments is prevented, and hence the Kondo scale $T_K$ vanishes 
at the quantum critical point (QCP). This scenario has, therefore, been termed 
local quantum critical (LQC) \cite{qsi01,coleman01}
CeCu$_{6-x}$Au$_x$ is one of the best characterized HF 
compounds \cite{Loehneysen94,Loehneysen96a,Loehneysen98,
Stockert98,Loehneysen98a,stroka93,grube99} with a QPT 
between a fully Kondo-screened paramagnetic and an 
antiferromagnetically ordered phase at a critical Au concentration 
of $x_c=0.1$. Based on inelastic neutron scattering 
experiments it was suggested that the LQC scenario is realized in this 
compound \cite{schroeder00}, yet this issue has remained controversial.

In order to address the crucial question whether Kondo screening persists or
not at the QPT, we present in this
work ultraviolet photoemission spectroscopy (UPS) measurements, which provide 
the most immediate access to the screening scale $T_K$
by directly recording the Kondo resonance (KR) in the local Ce~$4f$ 
spectrum \cite{kondo_reinert01,ehm07}. 
By performing the experiments
at temperatures $T$ well above the Ne\'el temperature and above the
lattice coherence temperature ($T>\{T_K,\,T_{coh};\,T_N\}$)
we are able to investigate the local spin screening effect {\it alone},
without the complications of Fermi-volume change due to lattice coherence, 
which influence the results of other experimental techniques. 

The high-resolution UPS was performed with a Gammadata R4000 
analyzer and a monochromatized VUV-lamp at
$h\nu\!=\!40.8$~eV (see Refs. \cite{ehm07} and \cite{klein08} for more details). 
We cleaved the single crystalline samples {\em in situ} just before 
the measurement, already at the measurement temperature. 
The surface is known to be $\gamma$-Ce-like, meaning that the surface 
properties do not differ from the bulk properties \cite{ehm07}.
\begin{figure*}[tb]
  \begin{center}
    \includegraphics[width = 0.92\linewidth,clip]{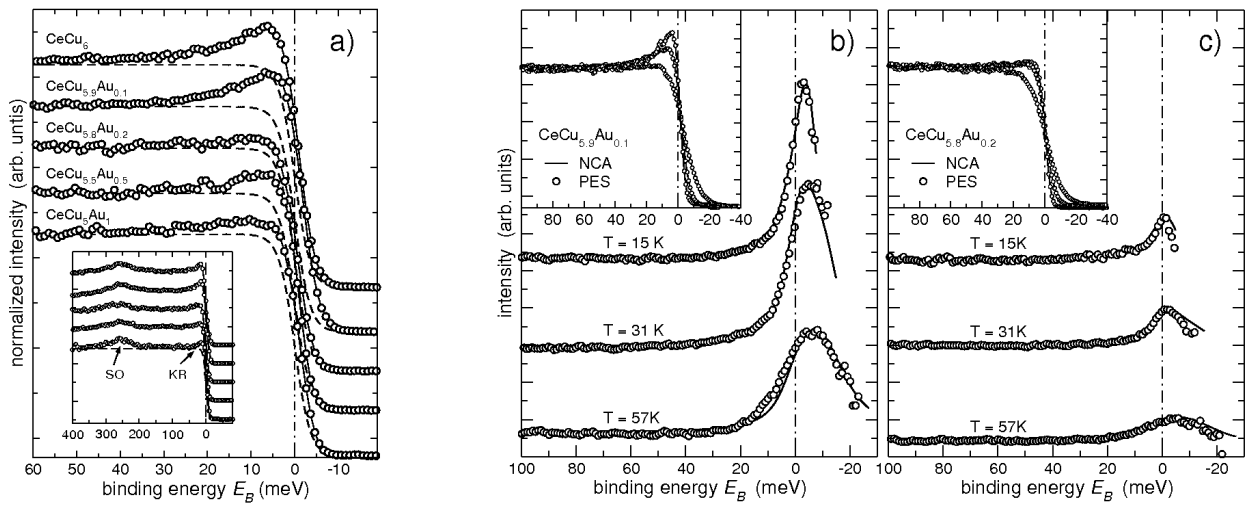}
    \caption[spectra]{
      {a)} Near-$E_F$ spectra of CeCu$_{6-x}$Au$_x$ for
      five different Au concentrations at $T\!=\!15$~K
      ($h\nu\!=\!40.8$~eVn $\Delta E\!=\!4.9$~meV), normalized 
      at $E_B\!\approx\!100$~meV. 
      The dashed lines describe the resolution broadened
      FDD at $T\!=\!15$\,K. The inset shows a larger energy 
      range including the spin-orbit (SO)
      partner at $E_B\!\approx\!260$~meV ($\Delta E\!=\!15$~meV)).  
      {b)} and {c)}  show spectra for $x\!=\!0.1$ and
      $x\!=\!0.2$, respectively, devided by the
      FDD, at various $T$. The solid lines are best NCA fits
      (see Tab.~\ref{tab:parameter} for the resulting model parameter values). 
      The insets in {b)} and {c)} show the corresponding raw data.
    \label{fig:spectra}}
  \end{center}
     \vspace*{-0.7cm}
\end{figure*}

The inset of Fig.~\ref{fig:spectra}{a)} shows survey spectra for 
five Au concentrations $x$ at $T\!=\!15$~K. 
The energy range includes the 
spin-orbit (SO) feature at 
$E_B\!\approx\!260$~meV and the tail of the Kondo resonance
just below the Fermi energy $E_F$. A distinct crystal-field (CF) feature
as observed,
e.g., in CeCu$_2$Si$_2$ and CeNi$_2$Ge$_2$ 
\cite{kondo_reinert01,ehm07}, 
is expected near $E_B=7$~meV, but strongly broadened 
\cite{stroka93} and, hence, not discernible in CeCu$_{6-x}$Au$_x$. 
The SO peak does not
show a significant $x$-dependent variation in position or intensity. 
This is expected, since it corresponds to an only virtually 
excited initial state of the UPS process \cite{kondo_reinert01,kroha03}.
In order to investigate the KR in more detail we performed 
high-resolution measurements near $E_F$ (Fig.~\ref{fig:spectra}{a)}). 
For all concentrations one observes a significant rise in the
intensity from $E_{B}\!\approx\!20$~meV towards $E_F$,
attributed to the tail of the KR which has its 
major spectral weight a few meV \textit{above} 
the Fermi energy \cite{kondo_reinert01, ehm07} 
suppressed by the Fermi-Dirac distribution function (FDD) in the UPS data.  
Normalizing to the FDD allows
to recover the thermally occupied spectrum up to $\approx 5k_BT$ above
$E_F$ \cite{kondo_reinert01, ehm07}
and reveals the KR. Fig.~\ref{fig:spectra} 
shows the FDD-renormalized spectra  
for $x\!=\!0.1$ ({b)})  and  $0.2$ ({c)}). 
In the low-$T$ spectra ($T\!=\!15$\,K) its maximum is at 
about $3$~meV above $E_F$ for $x\!=\!0.1$ and at $1$~meV for $x\!=\!0.2$.
The striking feature seen in Figs.~\ref{fig:spectra} {b)} and {c)}
is the significant drop of the spectral weight of the KR 
from $x\!=\!0.1$ to $0.2$, i.e., in the vicinity of the critical 
concentration $x_c$. 
The differences in the shape and the $T$ dependence between the  
$x\!=\!0$ and
$0.1$ spectra as well as the differences between the $x\!=\!0.2$, 
$0.5$,
and $1.0$ spectra (not shown) are negligibly small compared to the jump 
close to $x_c$. 
\begin{table}[b]
\vspace*{-0.4cm}
 \begin{tabular}{p{0.2cm} c p{0.5cm} c p{0.5cm} c p{0.5cm} c p{0.2cm}}
   & $x$ && $\Delta_{CF1, 2}$ (meV) && $V$ (meV) && $T_K$ (K)& \\ \hline
   & 0   && 7.2, 13.9 && 116 && 4.6& \\
   & 0.1 && 7.2, 13.9 && 116 && 4.6&\\ 
   & 0.2 && 8.7, 13.6 && 111 && 3.4& \\ 
   & 0.5 && 8.9, 13.6 && 109 && 3.3& \\
   & 1   && 9.2, 13.6 && 108 && 3.1& 
 \end{tabular} 
 \caption{SIAM/NCA fit parameters, CF splittings of the 
  Ce $4f$ orbitals, $\Delta_{CF1}$, $\Delta_{CF2}$, 
  $4f$-conduction band hybridization matrix element, $V$,  
  and the resulting Kondo temperatures, $T_K$, for different Au 
  concentrations $x$. Fixed parameters:
   conduction-band half-width $D_0\!=\!2.8$~eV (HWHM), single-particle 
   $4f$ binding energy
   $\varepsilon_f\!=\!-1.05$~eV, SO splitting $\Delta_{SO}\!=\!250$~meV. 
 \label{tab:parameter}}
\end{table}

\begin{figure}[t]
  \begin{center}
    \includegraphics[width = 0.92\linewidth,clip]{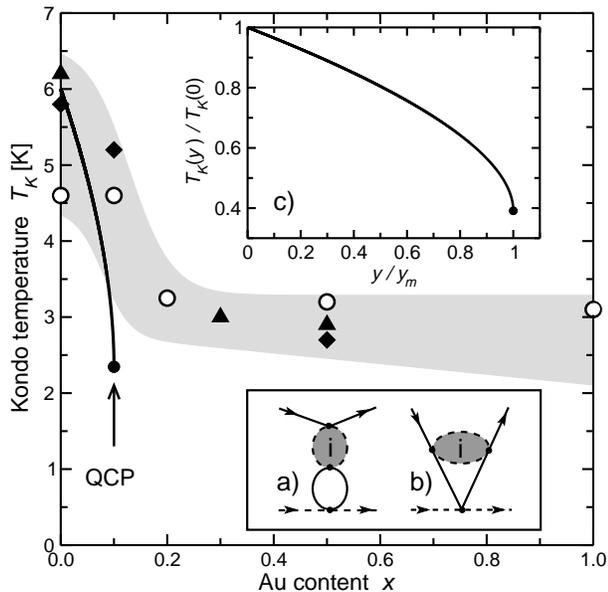}
    \caption[xdependence]{Dependence of the
      Kondo temperature $T_K$ on the Au content $x$, as determined by
      UPS (open circles), specific heat \cite{schlager92} 
      (triangles) and neutron scattering 
      \cite{stroka93,schroeder00}
      (diamonds).   
      The shaded area is a guide to the eye.
      The insets {a)} and {b)} show diagrammatic representations 
      of the RKKY
      corrections to the local Kondo vertex.
      The inset {c)} and the solid line in the main panel show the 
      universal curve $T_K(y)/T_K(0)$ vs. $y/y_{m}$ as given by 
      Eq.~(\ref{eq:TK_RG}).
    \label{fig:tkverlauf}}
  \end{center}
     \vspace*{-0.6cm}
\end{figure}

To extract the Kondo screening scale $T_K$ from the 
experimental results, one must bear in mind that the coherence 
temperature $T_{coh}$ obtained from resistivity measurements
\cite{loehneysen06} as well as the Kondo scale extracted from
specific heat \cite{schlager92} or neutron scattering data
\cite{stroka93,schroeder00}, are well below our lowest 
experimental temperature. Moreover, quantum critical fluctuations,
which in CeCu$_{6-x}$Au$_x$ extend up to about $T\!=\!7$~K 
\cite{schroeder00} and certainly become crucial at low $T$,
are not expected to influence our experimental spectra.
Thus, our data exhibit the onset 
of the {\it local\/} Kondo physics on the Ce atoms only.
Therefore, the single-impurity Anderson model (SIAM) is employed to 
interpret the experimental data. To determine $T_K$ we follow the 
procedure successfully applied to various Ce compounds 
in the past 
\cite{Patthey90,garnier97,allen00,kondo_reinert01,ehm07,ehm_sces03}:
Using the noncrossing approximation (NCA) %\cite{costi96, bickers87}
\cite{costi96} we calculate the Ce $4f$ spectral function of the SIAM,  
including all CF and SO excitations. For each composition $x$ the NCA 
spectra are broadened by the experimental resulotion and fitted to the 
experimental data, using a single parameter set for all experimental $T$.
The NCA spectra are then computationally extrapolated to   
$T\!\approx\!0.1\,T_K$, where $T_K$ is extracted from the 
Kondo-peak half-width at half maximum (HWHM) of the NCA spectra.   
%starting with parameters known from an earlier anaysis of the host compound 
%CeCu$_6$ \cite{ehm_sces03}.
The resulting fit parameter values are shown in Tab.~\ref{tab:parameter}.
Fig.~\ref{fig:tkverlauf} shows the $x$ dependence of $T_K$ for
CeCu$_{6-x}$Au$_x$ obtained from our UPS data,
compared to results of various other experiments \cite{schlager92,
stroka93,loehneysen06}. We emphasize that,
irrespective of a possible systematic ambiguity in the fit procedure,
the surprisingly abrupt step of $T_K$ near $x\!=\!x_c\!\approx\!0.1$ is
significant and already clearly visible in the raw 
data (Fig.~\ref{fig:spectra}).\\ \indent
What can be learnt from this step-like behavior at elevated 
$T$ about the nature of the QPT at $T\!=\!0$~K? 
The method of extracting a Kondo scale from 
a single-ion picture described above mimics exactly a 
system of dense Kondo atoms that, however, would {\it not} undergo any 
lattice coherence or magnetic ordering at sufficiently low $T$. 
This scenario is different from the low-$T$ behavior of  
the Kondo lattice and two-impurity Kondo problems, where Kondo spin
screnning would have to compete with the formation of inter-impurity  
coherent states \cite{varma87} or could be destroyed
by critical fluctuations of a magnetic order parameter 
\cite{varma05}. By contrast, our system is represented 
by an effective single-impurity Kondo model where the Kondo exchange 
coupling $J$ is renormalized by the local spin fluctuations on the surrounding 
{\it identical} Kondo atoms through the indirect Ruderman-Kittel-Kasuya-Yosida
(RKKY) coupling. This problem can be treated in a controlled way 
by renormalization group (RG) enhanced perturbation theory which allows for
definite predictions about the formation of HF quasiparticles at low 
$T$, as we will show in the following. 

\begin{figure}[t]
  \begin{center}
    \includegraphics[width = 0.92\linewidth,clip]{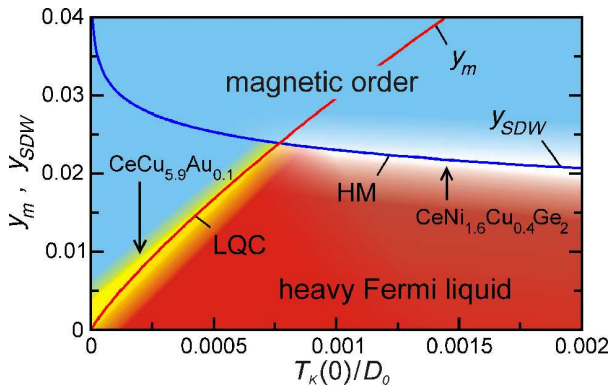}
    \caption[phasediagram]{(color online)
     Schematic $T\!=\!0$ phase diagram of a HF system with
     a magnetic QPT driven by RKKY coupling or a SDW instability 
     in the plane of single-ion
     Kondo-scale $T_K(0)/D_0$ and RKKY coupling strength $y$, 
     as drawn from the high-$T$ analysis.    
     Red line ($y_m$): maximum RKKY coupling strength beyond which  
     complete Kondo screening ceases to exist, Eq.~(\ref{eq:ymax}). 
     At the red line a step occurs in $T_K(y)$. 
     %as extracted from high-temperature spectral data. 
     Blue line ($y_{\text{SDW}}$): RKKY coupling strength at which 
     the heavy Fermi liquid becomes unstable towards a SDW, assuming a
     spin-spin coupling $J_{SDW}=J$. 
     The cases 
     $y_m < y_{\text{SDW}}$ and  $y_m > y_{\text{SDW}}$ distinguish
     whether the QPT is LQC-like (yellow)  or HM-like (white), 
     respectively. 
     The arrows indicate estimates \cite{ehm07} for
     $T_K(0)/D_0$ for CeCu$_{6-x}$Au$_x$
     and for CeNi$_{2-x}$Cu$_x$Ge$_{2}$. 
      \label{fig:phasediagram}}
      \end{center}
     \vspace*{-0.7cm}
\end{figure}

The leading perturbative corrections to $J$ on
a given Ce atom at lattice site $i=0$ are depicted diagrammatically in the
insets {a)}, {b)} of Fig.~\ref{fig:tkverlauf}. They obviously involve the full 
dynamical spin susceptibility $\chi_{4f}(T,\Omega)$ of a local Ce {\it 4f} 
spin on a neighboring site $i\neq 0$, where 
$\chi_{4f}(T,0)\!=\!(g_L\mu_B)^2\,N(0) D_0/(4\sqrt{T_K^2+T^2})$,
with $N(0)$ the density of states at the Fermi level, $D_0\approx E_F$ the
band cutoff and
$g_L$, $\mu_B$ the Land\'e factor and the Bohr magneton, respectively. 
Summing over all lattice sites $i\!\neq\!0$, one obtains for the  direct 
and the exchange corrections (Fig.~\ref{fig:tkverlauf} 
{a)} and {b)}), 
respectively, the spin-isotropic terms,
\begin{eqnarray}
\delta J^{(d)} &=& - y \frac{1}{4} J g_{i}^2 \ \frac{D_0}{\sqrt{T_K^2+T^2}}\
\frac{1}{1+(D/T_K)^2}
\label{eq:deltaJ_d}\\
\delta J^{(ex)} &=& - y \frac{1}{4} J g_{i}^2 
\left( \frac{3}{4} + \frac{T}{\sqrt{T_K^2 + T^2}} \right)\ .
\label{eq:deltaJ_ex}
\end{eqnarray}
Here $J$ is the bare, local spin exchange coupling on site $i\!=\!0$, 
and $g_i\!=\!N(0)J_i$ is the dimensionless, bare coupling on site $i\!\neq\!0$.
$y$ is a dimensionless factor that describes the 
(experimentally not known)
relation between the RKKY coupling strength and the Au content $x$. 
We assume a linear dependence across the QPT, $y=\alpha (x+x_0)$,
with adjustable parameters $\alpha$ and $x_0$.
As seen from Fig.~\ref{fig:tkverlauf} {a)}, $\chi_{4f}(T,\Omega)$ imposes a
soft cutoff on the energy exchange in $\delta J^{(d)}$, which appears as the
last factor in Eq.~(\ref{eq:deltaJ_d}) in terms of the running 
band cutoff $D\leq D_0$.  
In $\delta J^{(ex)}$ (Fig.~\ref{fig:tkverlauf} {b)}) 
there is no such cutoff; 
however, $\chi_{4f}(T,\Omega)$ restricts the conduction electron
response to the Fermi energy, and suppresses $\delta J^{(ex)}$ compared to
$\delta J^{(d)}$ by an overall factor of $\sqrt{T_K^2+T^2}/D_0$, as seen in 
Eq.~(\ref{eq:deltaJ_ex}). The one-loop RG equation for the 
local spin exchange coupling, including RKKY corrections, 
Eqs.~(\ref{eq:deltaJ_d}) and (\ref{eq:deltaJ_ex}), reads, 
%\vspace*{-0.2cm}
%\begin{eqnarray}
${d J}/{d\ln D}\!=\!-2 N(0) 
\left[ J+\delta J^{(d)}+\delta J^{(ex)} \right]^2$.
%\label{eq:RGequation}
%\end{eqnarray}
Note that in this RG equation the couplings $g_i$ on sites $i\!\neq\!0$ 
are not renormalized, 
since this is already included in the full susceptibility $\chi_{4f}$. 
The essential feature is that in the low-$T$ limit ($T\ll T_K$), 
for which $T_K$ is to be extracted, the direct RKKY correction
$\delta J^{(d)}$, Eq.~(\ref{eq:deltaJ_d}), is inversely proportional to the 
renormalized Kondo scale $T_K(y)$ itself via $\chi_{4f}(0,0)$.
$T_K(y)$, including perturbative RKKY corrections, is
defined as the scale where the solution of the RG equation diverges. 
This leads without further approximations to a highly non-linear 
renormalization of $T_K$ given by the selfconsistency equation,
\begin{eqnarray}
\frac{T_K(y)}{T_K(0)} = \exp
\left\{-\left(\frac{1}{2g} +\ln 2\right) \
\frac{f(u)}{1-f(u)}
\right\}  \ ,
\label{eq:TK_RG}
\end{eqnarray}
with $g\!=\!N(0)J$, $f(u)\!=\!u-u^2/2$, $u\!=\!yg^2D_0/[4T_K(y)]$.
The single-ion Kondo scale without RKKY coupling is
$T_K(0)=D_0\ {\rm exp}[-1/2g]$.
We have verified that this perturbative RG treatment of the RKKY corrections
is controlled in the sense that the effective perturbation parameter 
$f(u)\!\leq\!0.1$ for all selfconsistent solutions, i.e., the exponent  
in Eq.~(\ref{eq:TK_RG}) remains small.
The solution of Eq.~(\ref{eq:TK_RG}) is universal in the
dimensionless variables $T_K(y)/T_K(0)$ and $y/y_{m}$ and turns out to exist 
only for $y$ smaller than a maximum value $y_{m}$. For $y/y_{m}\to 1$ 
it behaves as $T_K(y)/T_K(0)\!=\!r + b \sqrt{1- y/y_{m}}$, with
numerically determined constants $r\equiv T_K(y_m)/T_K(0)\!\approx\!0.391$, 
$b\!\approx\!0.517$.
(cf. Fig.~\ref{fig:tkverlauf}{c)}).
The maximum RKKY parameter $y_{m}$ beyond which a solution of 
Eq.~(\ref{eq:TK_RG}) ceases to exist, can be deduced 
%as the point where the derivative $dT_K(y)/dy$ diverges. It reads
in terms of $\tau_K=T_K(0)/D_0$ as,
\begin{eqnarray}
y_{m}=8\,r\,\tau_K\,(\ln \tau_K)^2\hspace{-0.2ex}
\left[2\hspace{-0.2ex}-\hspace{-0.2ex}\ln\frac{\tau_K}{2}\hspace{-0.2ex}-
\hspace{-0.2ex}\sqrt{\left(2\hspace{-0.2ex}-\hspace{-0.2ex}
\ln\frac{\tau_K}{2}\right)^2\hspace{-0.2ex}-4} 
\right]
\label{eq:ymax}
\end{eqnarray}
(Fig.~\ref{fig:phasediagram}, red line).
For $y\!>\!y_{m}$, the Kondo spin exchange coupling including 
RKKY corrections does not diverge under RG. Hence, 
as the essential result of this analysis, in a dense system of Kondo ions,
complete Kondo screening ceases to exist above a critical RKKY coupling
strength, $y\!>\!y_{m}$, even if magnetic ordering does not occur. 
For comparison, the well-known Doniach criterion \cite{doniach77}
pertains to the breakdown of Kondo screening due to magnetic ordering and 
reads $T_K(0)\approx y_m N(0)J^2$.
It is similar, albeit not equivalent, in numerical values to 
Eq.~(\ref{eq:ymax}) and does not allow for a determination of the 
composition-dependent screening scale $T_K(y)$.
By identifying $y_{m}$ with the QCP, as done in Fig.~\ref{fig:tkverlauf},
the $x$ dependence of $T_K[y(x)]$ near this critical endpoint explains  
the abrupt step of $T_K$ observed experimentally 
in the high-$T$ spectral data.
The small KR spectral weight seen for $x\!>\!x_c$ should 
be interpreted merely as the logarithmic onset of Kondo screening
which does not fully develop even for $T\to 0$. Consequently, 
the small, residual local moments must order in dimensions $d>2$ at 
sufficiently low $T$.

Hence, the theory predicts two possible scenarios, depending on the 
size of $T_K(0)/D_0$: 
(1) Magnetic ordering at $T=0$ occurs for an RKKY parameter $y=y_{\text{SDW}}<y_m$,
i.e., without breakdown of Kondo screening. The ordering  may arise 
in this case from a $T=0$ spin-density-wave (SDW) instability of the underlying
heavy Fermi liquid. This corresponds to the HM scenario, depicted in 
Fig.~\ref{fig:phasediagram} as the white region. 
In this case, a step-like behavior of 
$T_K$ as extracted from the high-$T$ UPS spectra is still predicted 
from Eq.~(\ref{eq:TK_RG}) at $y=y_m$, i.e., inside the region where 
magnetic ordering occurs at $T<T_N$. 
(2) Magnetic ordering does not occur for $y<y_m$. 
In this case Eq.~(\ref{eq:TK_RG}) indicates a breakdown of Kondo screening 
at the magnetic QCP, and quantum critical fluctuations (not considered in 
our theory) will suppress the low-$T$ spin screening scale to zero 
at this point. This is the LQC scenario, shown in 
Fig.~\ref{fig:phasediagram} as the yellow region.\\ \indent  
To conclude, the theoretical analysis predicts generally 
that an abrupt step of the Kondo 
screening scale extracted from high-$T$ spectral data
should occur in any HF compound with competing Kondo and RKKY
interactions, as long as the single-ion Kondo screening scale is 
larger than the magnetic ordering temperature. Whether this
distinct feature is located at the quantum critical control parameter
value $x_c$ or inside the magnetically ordered region constitutes a general
high-$T$ criterion to distinguish the LQC and HM scenarios.
The sharp step of $T_K$ occurring in our UPS data
very close to $x_c$ (see Fig.~\ref{fig:tkverlauf}) 
strongly supports that CeCu$_{6-x}$Au$_x$ 
falls into this class, as was previously inferred
indirectly from inelastic neutron scattering data 
\cite{schroeder00}.\\ \indent
We would like to thank F. Assaad, S. Kirchner, and M. Vojta
for fruitful discussions.
This work was supported by DFG through
Re~1469/4-3/4 (M.K., A.N., F.R.), SFB~608 (J.K.) and FOR~960 (H.v.L.).

%\vspace*{-0.3cm}

\end{document}